%% file: main.tex
\documentclass[runningheads]{llncs}

\input{preamble/packages}

\begin{document}
\title{Tutorial on\\Reasoning for IR \& IR for Reasoning}
\author{
  Mohanna Hoveyda\inst{1} \and
  Panagiotis Eustratiadis\inst{2} \and \\
  Arjen P. de Vries\inst{1} \and
  Maarten de Rijke\inst{2}
}

\institute{
  Radboud University \and University of Amsterdam\\
\email{
mohanna.hoveyda@ru.nl, p.efstratiadis@uva.nl,\\
arjen.devries@ru.nl,
m.derijke@uva.nl
}  
}

\authorrunning{Hoveyda et al.}

\maketitle             

\begin{abstract}
Information retrieval has long focused on ranking documents by semantic relatedness. Yet many real-world information needs demand more: enforcement of logical constraints, multi-step inference, and synthesis of multiple pieces of evidence. Addressing these requirements is, at its core, a problem of \textit{reasoning}. 
Across AI communities, researchers are developing diverse solutions for the problem of reasoning, from inference-time strategies and post-training of LLMs, to neuro-symbolic systems, Bayesian and probabilistic frameworks, geometric representations, and energy-based models. 
These efforts target the same problem: to move beyond pattern-matching systems toward structured, verifiable inference. However, they remain scattered across disciplines, making it difficult for IR researchers to identify the most relevant ideas and opportunities.
To help navigate the fragmented landscape of research in reasoning, this tutorial first articulates a working definition of reasoning within the context of information retrieval and derives from it a unified analytical framework. The framework maps existing approaches along axes that reflect the core components of the definition.
By providing a comprehensive overview of recent approaches and mapping current methods onto the defined axes, we expose their trade-offs and complementarities, highlight where IR can benefit from cross-disciplinary advances, and illustrate how retrieval process itself can play a central role in broader reasoning systems. The tutorial will equip participants with both a conceptual framework and practical guidance for enhancing reasoning-capable IR systems, while situating IR as a domain that both benefits and contributes to the broader development of reasoning methodologies.
\end{abstract}

\input{sections/01-motivation}
\input{sections/02-learning-objectives}
\input{sections/03-scope}
\input{sections/04-relevance}
\input{sections/05-format}

\input{sections/07-detailed-outline}
\input{sections/08-target-audience}
\input{sections/09-history}
\input{sections/10-presenters}
\input{sections/11-materials}
\vspace*{-2mm}
\vspace*{-2mm}
\vspace*{-2mm}
\subsubsection*{Acknowledgements.}
This research was (partially) supported by the Dutch Research Council (NWO), under project numbers 024.004.022, NWA.1389.20.\-183, and KICH3.LTP.20.006, and the European Union under grant agreements No. 101070212 (FINDHR) and No. 101201510 (UNITE).
All content represents the opinion of the authors, not necessarily shared or endorsed by their respective employers and/or sponsors.
\vspace*{-2mm}
\vspace*{-2mm}
\subsubsection*{Disclosure of interests.}
\vspace*{-2mm}
The authors have no competing interests to declare that are relevant to the content of this article. 
\vspace*{-2mm}
\vspace*{-2mm}

\bibliographystyle{splncs04nat}
{\small
\bibliography{references-shortened}
}

\end{document}

%% file: preamble/packages.tex
\usepackage[T1]{fontenc}
\usepackage[inline]{enumitem}
\usepackage{graphicx}
\usepackage[numbers,square,sort&compress,sectionbib]{natbib}
\usepackage[colorlinks=true,
            linkcolor=blue,
            citecolor=blue,
            filecolor=blue,
            urlcolor=blue]{hyperref}
\usepackage{xcolor}

%% file: sections/01-motivation.tex
\vspace*{-2mm}
\vspace*{-2mm}
\section{Motivation}
\vspace*{-2mm}
While dense and generative models excel at semantic matching across queries and documents, real-world information needs often demand reasoning: enforcing negation \cite{petcu2025comprehensive,DBLP:conf/eacl/WellerLD24}, exclusion \cite{zhang2025excluir}, composing sets \cite{malaviya-etal-2023-quest}, and synthesizing (conflicting) evidence across various sources remain difficult for neural retrievers despite advances \cite{van_den_Elsen_2025}. Recent IR benchmarks contain queries that expose clear empirical weaknesses of current retrieval systems~\cite{killingback2025benchmarkinginformationretrievalmodels,DBLP:conf/iclr/SuYXSMWLSST0YA025,DBLP:journals/corr/abs-2508-06600}. 

In parallel, theoretical findings indicate that embedding-based retrieval methods, despite their empirical success, are subject to intrinsic representational limits, that may not be alleviated by pure scale of model and/or data size \cite{weller2025theoreticallimitationsembeddingbasedretrieval}. Large language models (LLMs), though often regarded as general-purpose reasoning systems and used as an alternative to transcend such limitations through generative reasoning, nonetheless exhibit persistent weaknesses in grounding and inference control, with outputs frequently lacking evidential support~\cite{kalai2025language}.
Considered together, these results suggest that information retrieval is reaching the limits of pattern-matching systems. In various subfields of AI, building models that can perform \textit{reasoning} has once again attracted significant attention. 

Inference-time strategies seek to elicit structured  `\,thinking\,' from LLMs, through chain-of-thought and iterative refinement \cite{DBLP:conf/nips/Wei0SBIXCLZ22,madaan23,yao2023tree}. Reinforcement learning (RL) has been explored as a mechanism to induce and regulate reasoning in LLMs, by defining reward signals over intermediate derivations, aligning model trajectories with human or automated preferences, and coupling language generation with structured search and planning procedures \cite{ouyang2022training,jin2025searchr}.
Neuro-symbolic approaches couple statistical generalization with formal inference \cite{Olausson_2023}. Probabilistic and Bayesian frameworks cast reasoning as inference under uncertainty \cite{DBLP:conf/emnlp/ParuchuriGLHSA024}. 
Some other line of work focuses on alternative representation spaces, using non-Euclidean embeddings to capture hierarchical structure \cite{DBLP:journals/taslp/ChenHLHXZLS24} or compositional representations to model Boolean operations in queries \cite{DBLP:conf/nlpir/MaiG024}. Lastly, energy-based formulations interpret reasoning as iterative optimization in latent space \cite{du24energydiff}.
These directions offer avenues with distinct strengths, but the approaches remain fragmented across disciplines. They highlight aspects of the reasoning problem, making it essential to connect them within a bigger picture of how representations, inference mechanisms, and learning signals interact.

IR both benefits from and contributes to this momentum. It provides a natural testbed where reasoning models must operate at scale, ground their steps in external evidence, and be judged by rigorous evaluation. The goal of this tutorial is to consolidate reasoning methodologies into a coherent framework for IR, and to show how IR can serve not only as a beneficiary but also as a useful ground for increasingly robust reasoning architectures.

%% file: sections/02-learning-objectives.tex
\vspace*{-2mm}
\section{Learning Objectives}
\vspace*{-2mm}
The tutorial is designed to provide a conceptual overview of and practical guidance into reasoning for IR. Specifically, participants will be able to:

\begin{enumerate}
    \item Identify the requirements of complex IR tasks, including negation, exclusion, set composition, temporal constraints, and multi-hop evidence synthesis, as well as an overview of applicable benchmarks.
    \item Understand the main methodological families for reasoning: \begin{enumerate*}[label=(\roman*)]
\item \textit{LLM-based (including inference-time strategies, and reinforcement learning approaches)}, \item \textit{neuro-symbolic integration}, \item \textit{probabilistic frameworks}, \item \textit{alternative representation spaces and optimization approaches}.
\end{enumerate*}
    \item Compare methods along three axes: \begin{enumerate*}[label=(\roman*)]
\item  \textit{representational adequacy}, \item \textit{mechanisms of inference and learning}, and \item \textit{computational viability for IR.} \end{enumerate*} They can consequently analyze trade-offs between approaches with respect to requirements of an IR system.
    \item Apply the conceptual framework introduced in the tutorial to propose novel solutions for IR and recognize opportunities where IR can enhance the development of more general reasoning architectures.
\end{enumerate}
\vspace*{-2mm}

%% file: sections/03-scope.tex
\vspace*{-2mm}
\vspace*{-2mm}
\section{Scope}
\vspace*{-2mm}
The area to be covered is broad and touches on many sub-discipliens in IR and AI. Since the topics covered are broad and technically diverse, the tutorial will not aim for exhaustive coverage of each reasoning method and its use or potential in or with IR, but rather to provide participants with an overarching view of the challenges and the research possibilities. Slides and reference materials will be shared with participants so as to help them to continue to explore the area.

%% file: sections/04-relevance.tex
\vspace*{-2mm}
\section{Relevance}
\vspace*{-2mm}
Recent IR benchmarks expose clear weaknesses of current models in handling complex user queries. Meanwhile, advances in LLM-based and neuro-symbolic, and other alternative approaches to reasoning remain scattered across fields, limiting their practical uptake in IR. This tutorial unifies these developments within a coherent analytical framework, clarifying how reasoning can be defined, modeled, and evaluated in IR and and helping researchers identify transferable methods and understand their trade-offs.

%% file: sections/05-format.tex
\vspace*{-2mm}
\vspace*{-2mm}
\section{Tutorial Format}
\vspace*{-2mm}
The format is primarily lecture-style, organized in four main blocks (see the detailed outline below), but will include interactive polls and short Q\&A slots after each block to encourage engagement.

%% file: sections/07-detailed-outline.tex
\vspace*{-2mm}
\vspace*{-2mm}
\section{Detailed Outline}
\vspace*{-2mm}
\vspace*{-2mm}
We will teach a half-day (3 hours) tutorial, aligned with the two breaks organized by the conference. The tutorial will be organized in four main blocks:
\vspace*{-2mm}
\vspace*{-2mm}

\subsection{Introduction (15 minutes)} 
\vspace*{-2mm}

Why reasoning is central to information retrieval; overview of tutorial's objectives, structure, and relevance to IR community. 
\vspace*{-2mm}
\subsection{Reasoning: Definition and its Challenges in IR (30 minutes)}
    \begin{enumerate}[label=(\arabic*)]
      \item Definition of reasoning in IR.
      \item Empirical and theoretical challenges:
        \begin{enumerate}
          \item \textit{Tasks and datasets:} NevIR~\cite{DBLP:conf/eacl/WellerLD24,petcu2025comprehensive,van_den_Elsen_2025}; ExcluIR~\cite{zhang2025excluir}; QUEST~\cite{malaviya-etal-2023-quest}; \\BRIGHT \cite{DBLP:conf/iclr/SuYXSMWLSST0YA025}; BrowseComp-Plus~\cite{DBLP:journals/corr/abs-2508-06600}; complex retrieval tasks~\cite{killingback2025benchmarkinginformationretrievalmodels}.
          \item \textit{Limits of current systems:} theoretical limits of embedding-based retrieval~\cite{weller2025theoreticallimitationsembeddingbasedretrieval}; taxonomy for negation~\cite{petcu2025comprehensive}; LLM hallucinations~\cite{kalai2025language}.
        \end{enumerate}
    \end{enumerate}
\subsection{Methodological Families (90 minutes)} 
    \begin{enumerate}[label=(\arabic*)]
        \item \textit{LLM inference-time strategies and optimization for reasoning}: 
        Approaches that elicit or refine reasoning at LLMs' decode/test time, including chain-of-thought prompting~\cite{DBLP:conf/nips/Wei0SBIXCLZ22}, IRCoT for multi-step questions~\cite{DBLP:conf/acl/TrivediBKS23}, Self-Refine for iterative self-feedback and revision~\cite{madaan23}, test-time adaptation/active fine-tuning~\cite{hbotter2024efficiently}, and implicit chain-of-thought without explicit prompts~\cite{10.5555/3737916.3740039}. Complementary results link these behaviors to procedural knowledge acquired during pretraining of LLMs~\cite{DBLP:conf/iclr/RuisMBKGLKRGB25}.

        \item \textit{LLMs + RL}: 
        Reinforcement learning (RL) has been increasingly explored as a means to incentivize and enhance reasoning in LLMs. Building on instruction-following models trained with human feedback~\cite{ouyang2022training}, recent systems such as \textit{DeepSeek-R1}~\cite{DBLP:journals/nature/GuoYZSWZXZMBZY025} and \textit{Search-R1}~\cite{jin2025searchr} use structured rewards and interaction with search or planning environments to promote stepwise reasoning. In parallel, complementary analyses question whether such RL-based fine-tuning genuinely improves reasoning ability beyond the pretrained model~\cite{yue2025does}, and examine the limits of language-based policy optimization~\cite{ramamurthy2022reinforcement}.
        
        \item \textit{Neuro-symbolic approaches:} Early efforts to ground IR in non-classical logic \cite{10.1145/24634.24635} offered a rigorous theoretical foundation but faced scalability and practical limitations.
        Recent work on reasoning explores use of symbolic solvers, provers and logic engines alongside neural components to enhance the reliability of inference in reasoning-intensive tasks \cite{Olausson_2023}. LINC use the LLM as a semantic parser from natural language to first-order logic, delegating deductive inference to external theorem provers~\cite{Olausson_2023}. Other approaches target reasoning under uncertainty by prompting LLMs to produce formal representations (e.g., code, probabilistic logic) that can be executed or queried~\cite{10.1609/aaai.v39i23.34674}, or by synthesizing task-specific probabilistic formulations on demand to support bayesian inference for open-world cognition~\cite{wong2025modelingopenworldcognitionondemand,wong2023wordmodelsworldmodels}.

        \item \textit{Probabilistic and Bayesian Frameworks}: 
        This line of research approaches reasoning by explicitly modeling uncertainty and integrating Bayesian inference to achieve a probabilistic reasoning system~\cite{qiu2025bayesianteachingenablesprobabilistic,hoffman2023training,feng2025birdtrustworthybayesianinference,yin2024reasoning}.
        
        \item \textit{Alternative representation spaces and optimization approaches:}
         Several works enhance the representation space for queries and documents to support explicit set/logic operations and hierarchical structure, including:
         BoxEmbeddings~\cite{chheda-etal-2021-box},
         Set-Compositional~\cite{krasakis2025constructingsetcompositionalnegatedrepresentations}, , and hyperbolic representation spaces~\cite{Dong_Jamnik_Liò_2025,DBLP:journals/taslp/ChenHLHXZLS24,yang2024hyperbolicfinetuninglargelanguage,yang2024enhancing}. Complementarily, some alternative approaches model reasoning as iterative inference in latent space via energy-based optimization~\cite{du24energydiff,gladstone2025energybasedtransformersscalablelearners}.
\end{enumerate}
\vspace*{-2mm}
\vspace*{-2mm}
\vspace*{-2mm}
\subsection{Bridging Current Methodologies \& Future Directions (45 minutes)} 
\vspace*{-2mm}
We first introduce our comparative framework and then map the discussed methods regarding to the framework's axes;
\begin{enumerate}[label=(\arabic*)]
        \item \textit{Representational adequacy}; to what extent the representation space encodes logical structures, hierarchies and uncertainty about predicates across the various approaches,
        \item \textit{Mechanisms of inference verification and learning}; how inference and updates are applied over the representation space,
        \item \textit{Computational viability for IR}; to what extent each of the reasoning frameworks can be applied in a high-scale real-world IR setting.
    \end{enumerate}
We end with a roadmap of open problems and research directions for reasoning in IR and leveraging IR within broader reasoning models.

%% file: sections/08-target-audience.tex
\vspace*{-2mm}
\section{Target Audience}
\vspace*{-2mm}
The tutorial is designed for a broad audience within and beyond the IR community, with a focus on three groups: 
\begin{enumerate*}[label=(\roman*)]
\item \textit{Early-career IR researchers and PhD students} who wish to tackle complex information needs (e.g., negation, set-composition, and multi-hop), and who are seeking conceptual frameworks to guide new research directions.
    \item \textit{Applied researchers and practitioners} in search, recommendation, or conversational systems who want to enhance the robustness of their products in edge cases where current semantic-matching approaches fail. 
    \item \textit{Researchers from related fields} (e.g., NLP, ML, knowledge representation) who are interested in how diverse reasoning methodologies can be applied to IR and, conversely, how IR can serve general reasoning models. 
\end{enumerate*}
We assume familiarity with modern IR models, but no prior expertise in formal logic or specialized representations/optimization approaches, remaining accessible to newcomers yet technically engaging for experts.

%% file: sections/09-history.tex
\vspace*{-2mm}
\section{Tutorial History and Related Events}
\vspace*{-2mm}
This tutorial is new and prepared exclusively for ECIR 2026. Relevant past events with a reasoning theme across various venues in IR, NLP, and AI include:
\begin{enumerate}[label=(\arabic*)]
    \item EMNLP’20: Machine Reasoning: Technology, Dilemma and Future~\cite{duan-etal-2020-machine}
    \item ECIR’23: Neuro-Symbolic Approaches for Information Retrieval~\cite{10.1007/978-3-031-28241-6_33}
    \item ACL’23: Complex Reasoning in Natural Language~\cite{zhao-etal-2023-complex}
    \item AAAI’23: Advances in Neuro-Symbolic Reasoning and Learning~\cite{ShakarianSimariBaralVelasquez2023}
    \item KDD’23: Knowledge Graph Reasoning and Its Applications~\cite{10.1145/3580305.3599564} 
\end{enumerate}
Alongside the enumerated tutorials, several existing surveys on reasoning in LLMs, RL, optimization-based formulations, geometric representations, and logic-based IR will inform the content of this tutorial \cite{DBLP:journals/tmlr/Carbone25,10.1145/3711896.3736564,li2025implicitreasoninglargelanguage,DBLP:journals/tmlr/CasperDSGSRFKLF23,DBLP:journals/csur/AbdulahhadBCP19,hase2024fundamental}. In contrast to prior tutorials, we define reasoning specifically within the context of IR and organize recent methodological advances into a unified analytical framework which connects developments across representation, inference and learning, and computational viability, providing a clear structure for analyzing, comparing, and developing reasoning-capable information access systems.
\vspace*{-2mm}

%% file: sections/10-presenters.tex
\vspace*{-2mm}
\section{Presenters}
\vspace*{-2mm}
\textbf{Mohanna Hoveyda (Radboud University, primary contact)} is a PhD candidate in Conversational AI and IR. 
She develops frameworks for orchestrating multi-agent, LLM-based pipelines via reinforcement learning, and neuro-symbolic and probabilistic frameworks for complex information needs. Mohanna has previously taught classes in machine learning for natural language processing, artificial intelligence, and information retrieval.

\smallskip\noindent\textbf{Panagiotis Eustratiadis (University of Amsterdam)} is a postdoctoral researcher with expertise in matters of AI safety. His research focuses on adversarial robustness, Bayesian uncertainty estimation, and information retrieval. 
Panagiotis has taught diverse topics in machine learning and information retrieval at a BSc and MSc level, as well as AI safety-related topics at a PhD level. 

\smallskip\noindent\textbf{Arjen P. de Vries (Radboud University)} is professor of IR. His research aims to help understand how users and systems may cooperate to improve information access, and focuses on questions regarding the combination of structured and unstructured information representations. Arjen has taught at all levels on a broad range of topics in information retrieval and data science. He has co-taught a number of tutorials at ECIR, SIGIR and ESSIR.

\smallskip\noindent\textbf{Maarten de Rijke (University of Amsterdam)} is a Distinguished University Professor of Artificial Intelligence and Information Retrieval at the University of Amsterdam. 
Maarten has taught numerous courses at the BSc, MSc, and PhD level at the University of Amsterdam, numerous tutorials on a range of information retrieval topics at all the main conferences and summer schools in our field, and has co-organized multiple summer schools in information retrieval and artificial intelligence, most recently the 2025 ICAI Summer School.

%% file: sections/11-materials.tex
\vspace*{-2mm}
\vspace*{-2mm}
\section{Materials}
\vspace*{-2mm}
\vspace*{-2mm}
A dedicated website will contain slides, references, as well as pointers to libraries and experimental environments that help participants of our tutorial examine the area further. See
  \url{https://reasoning-for-ir.github.io}.